\documentclass[aps,pre,noeprint,superscriptaddress,amssymb]{revtex4-2} 
\usepackage{amsmath}
\usepackage{amssymb}
\usepackage{graphicx}
\usepackage{tabularx}
\usepackage{gensymb}
\usepackage{url,hyperref}
\usepackage{mathrsfs, float}
\usepackage{physics}

\usepackage[usenames,dvipsnames]{xcolor}
\hypersetup{colorlinks=true, linkcolor=BrickRed, urlcolor=blue!50!black, citecolor=blue!50!black}
\usepackage[capitalize]{cleveref}
\usepackage{cleveref}
\AtBeginDocument{}
\crefrangelabelformat{equation}{(#3#1#4)$-$(#5#2#6)}

\begin{document}

\title{Giant Diffusion at Flocking Transitions}
\author{Sohini Chatterjee}
\author{Wasim Akram}
\author{Subir K. Das}
\email{Email of the corresponding author: das@jncasr.ac.in}
\affiliation{Theoretical Sciences Unit and School of Advanced Materials, Jawaharlal Nehru Centre for Advanced Scientific Research, Jakkur P.O., Bangalore 560064, India}

\date{\today}

\begin{abstract}

We simulate dynamics in two-dimensional aligning active-matter models exhibiting order-disorder transitions in the velocity field. Analysis reveals that the single-particle diffusivity, $D_s$, exhibits strong enhancements at criticality which we quantify via a finite-size scaling method. This dynamical behavior at nonequilibrium flocking transitions is fundamentally different from that at the criticality of equilibrium systems, where diffusivities either vanish or remain non-critical. Furthermore, $D_s$ exhibits pronounced finite-size effects throughout the ordered phase, which also are absent in equilibrium settings. We interpret these counterintuitive transport anomalies within the framework of a flocking theory. Our results suggest that enhanced diffusion generates effective long-range interactions. This relaxes the constraints of the Mermin–Wagner theorem that otherwise prevents order-disorder transitions in two dimensions for systems possessing continuous symmetry. These results are of much experimental relevance and have important implications in formulating quantitative theories for nonequilibrium transports.

\end{abstract}

\maketitle

Slow dynamics is a hallmark of glassy behavior \cite{debenedetti2001} and critical phenomena \cite{fisherRevModPhys.46.597,hohenberg1977theory,tauberBook2014critical,onukiBook2002}, where diffusivities vanish and viscosities diverge. At a glass transition, microscopic self-diffusivity, $D_s$, tends to vanish \cite{debenedetti2001}. At the critical point of an equilibrium fluid, on the other hand, collective diffusivities vanish, complementing the divergences of viscosities, leading to critical slowing down \cite{hohenberg1977theory,tauberBook2014critical,burstyn_sengers_1980dynamical,sengers1982universality,burstyn_sengers_1983dynamic,hao_ferrell_bhatt_2005critical,bhattacharjee1982critical,chen2005universal,das2006critical,Roy2013JCP,Roy2014JCP,midya2017finite,shrivastav2025JCP}. In this case, $D_s$ remains insensitive to the critical fluctuations \cite{das2006critical,shrivastav2025JCP}. These dynamical features are universal in passive fluids.
Here we study transport phenomena on both sides of criticality of two distinct active-matter models \cite{vicsek1995novel,das2017pattern,Paul2024} that realistically capture flocking transitions for self-propelling agents. Our calculations reveal that $D_s$ exhibits strong critical \textit{enhancement}, which is in stark contrast to the equilibrium expectations \cite{hohenberg1977theory,tauberBook2014critical,onukiBook2002,stanleyBook,burstyn_sengers_1980dynamical}. 

In active matter \cite{Marchetti_rev,Ramaswamy2010Active,Bechinger_rev}, transitions \cite{vicsek1995novel,toner1995,toner1998flocks,czirok1997spontaneously,das2014phase,Trefz2016Activity,siebert2018critical,partridge2019critical,Dittrich2021Critical,Midya2023,albano2008finite,albano2012criticality,gregoire2004onset} are driven not by interparticle potentials but by dynamical rules \cite{vicsek1995novel,toner1995,das2017pattern,Paul2024,siebert2018critical,partridge2019critical,Midya2023,Wilding2021,Wilding2021_2,Julia2016,Julia2013,Julia2018,Cates2014,Cates2015,Cates2015prl,Lowen2019,Prabal2023,Lowen2025,Cugliandolo2024,Cugliandolo2020,Cates2013,Lowen2015,Jhajhria2025}. Existing studies of dynamics near the critical point, in such systems, nevertheless, have been limited only to estimating \cite{hohenberg1977theory,tauberBook2014critical,burstyn_sengers_1980dynamical,partridge2019critical,Midya2023,albano2008finite} the relaxation exponent $z$: $\tau\sim\xi^z$, $\xi$ and $\tau$ being, respectively, the correlation length and relaxation time \cite{fisherRevModPhys.46.597,hohenberg1977theory}. Given the distinct driving mechanisms underlying these nonequilibrium transitions, a proper understanding of dynamical critical behavior requires systematic studies of transport properties. Available transport related investigations \cite{Karmakar2025,Ryabov2022SoftMatter,Ignacio2019,wu2000particle}, however, are not specific to critical behavior.
In this study of two-dimensional systems, in addition to the critical enhancement, we also observe strong finite-size effects in $D_s$ throughout the ordered regions, accompanied by pronounced points of inflection. These imply a breakdown of the conventional picture of critical slowing down and suggest a violation of the Stokes-Einstein-Sutherland \cite{hohenberg1977theory,tauberBook2014critical,burstyn_sengers_1980dynamical,sengers1982universality,burstyn_sengers_1983dynamic,hao_ferrell_bhatt_2005critical,bhattacharjee1982critical,chen2005universal,das2006critical,Roy2013JCP,Roy2014JCP,midya2017finite,shrivastav2025JCP} relation in aligning active-matter. The results have important implications in the context of Mermim-Wagner (MW) theorem. We interpret the findings within the framework of a theory \cite{toner1995,toner1998flocks} on flocking transition. Given that the considered models have wide applicability --- in systems ranging from schools of fish to bacterial colonies to swarm robotics \cite{Marchetti_rev,Ramaswamy2010Active,Poschel2018, robot2024, holubec2025,robot_hybrid} --- our results are of broad fundamental and practical significance.

We study the Vicsek model \cite{vicsek1995novel}, to be referred to as model 1 (M1), with the dynamical rules: ${\rm{a)}}~~~ \vec{r}_j (t+\Delta t) = \vec{r}_j (t) + \vec{v}_j (t) \Delta t$, 
$\vec{r}_j (t)$ being the position of particle $j$ at time $t$, $\Delta t$ a discrete time step and $\vec{v}_j = v_0 e^{i \theta_j}$; ${\rm{b)}}~~~\theta_j (t+\Delta t) = \langle \theta (t) \rangle ^j_{ \mathcal{R}_{\rm{M1}} } + \Delta\theta$,
where the velocity-aligning term, $\langle \theta (t) \rangle ^j _{ \mathcal{R}_{\rm{M1}} }$,
represents the average direction of motion of particles in the neighborhood of a particle $j$, defined by a square cell of linear dimension $\mathcal{R}_{\rm{M1}}$. Through $\Delta\theta$, noise is introduced via uniform random numbers $\in[0,\eta]$. An order-disorder transition, for a system of $N$ particles, characterized by the order parameter 
    $v_a = [(1/N) | \sum_{j=1}^{N} e^{i \theta_j}|]$,
arises from a competition between $\eta$ and $\langle \theta (t) \rangle^j _{ \mathcal{R}_{\rm{M1}} }$. The transition point $(\eta_c)$ depends \cite{albano2008finite,albano2012criticality} upon the size of $\mathcal{R}_{\rm{M1}}$, the density ($\rho$) of particles and $v_0$, the (constant) magnitude of particle velocities. We perform simulations, with $\mathcal{R}_{\rm{M1}}=1$, $\rho=2.0$, $v_0=0.1$ and $\Delta t=1$, in periodic square boxes. For these choices, $\eta_c \simeq 2.8$ \cite{czirok1997spontaneously}.  
The results of M1 are nicely complimented by those from a second model (M2) \cite{das2017pattern,Paul2024,Paul2021}. In this realistic framework, with overlap preventing passive inter-particle interaction $U_i$, an active force $\vec{f}_i$ and a Langevin thermostat \cite{das2014phase,Trefz2016Activity,FrenkelSmit2023,allen2017}, for the $i^{\rm{th}}$ particle position $\vec{r}_i$ we solve \cite{das2017pattern,Paul2024,FrenkelSmit2023,allen2017}: $m d^2\vec{r}_i/dt^2 = -\gamma m d\vec{r}_i/dt - \vec{\nabla} U_i+ \sqrt{2m\gamma k_BT} \vec{\xi}_i(t)+\vec{f}_i$, with $m$, $\gamma$, $k_B$ and $T$ being mass, drag coefficient, Boltzmann constant and temperature, respectively.
The Gaussian white noise $\vec{\xi}_i$ has the correlation $\langle \xi_{i\mu}(t)\xi_{j\nu}(t') \rangle=\delta_{ij}\delta_{\mu\nu}\delta(t-t')$. For $U_i$, while the choice can be reasonably arbitrary, we chose a modified \cite{das2006critical} Lennard-Jones potential: $U(r)=u(r)-u(\mathcal{R_{\rm{M2}}})-(r-\mathcal{R_{\rm{M2}}}) \,dU/dr|_{r=\mathcal{R}_{\rm{M2}}}$, with $u=4\epsilon[(\sigma/r)^{12}-(\sigma/r)^{6}]$, $\epsilon$ being the interaction strength and $\mathcal{R}_{\rm{M2}}=2.5\sigma$, $\sigma$ representing the particle diameter.   
We adopt a Vicsek-like active force: $\vec{f}_i=f_A\hat{n}_i$, $\hat{n}_i(={\sum_j \vec{v}_j}/{|\sum_j \vec{v}_j|})$ being the average direction of neighbors within radius $\mathcal{R}_{\rm{M2}}$.

\begin{figure}
\centering
\includegraphics*[width=0.48\textwidth]{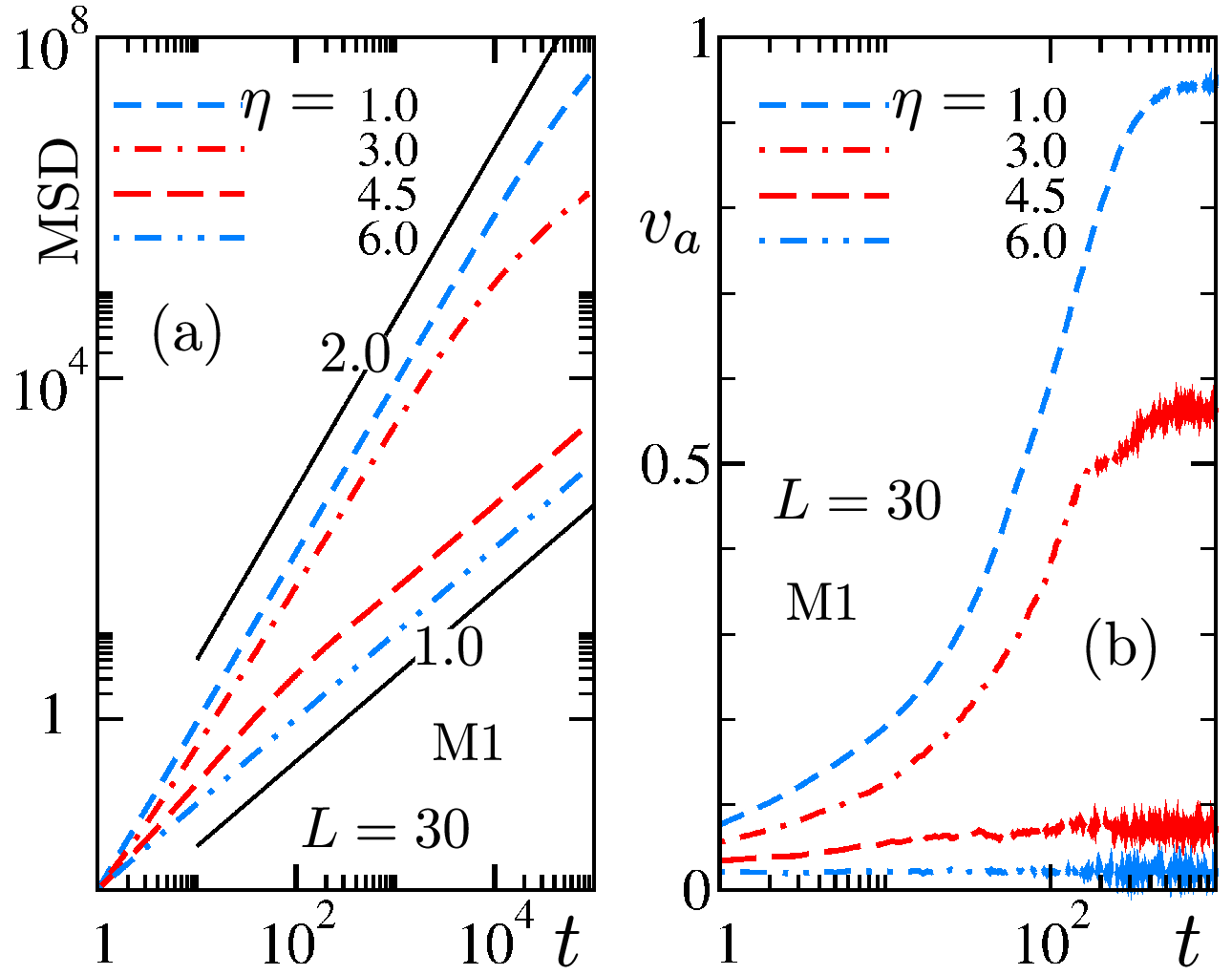}
\caption{(a) Mean squared displacement (MSD) is plotted as a function of time ($t$). Results are included from simulations with a few different values of the noise parameter $\eta$. The solid lines, with power-law exponents $1$ and $2$, represent diffusive and ballistic behavior, respectively. (b) The evolutions of the order parameter ($v_a$) are shown as a function of $t$, for the same set of $\eta$ values as in (a). These results are for M1.}
\label{fig_msd1}
\end{figure}

In Fig. \ref{fig_msd1} (a), for the M1, we show the average mean squared displacements (MSD) of the particles \cite{hansen2013theory}, as a function of $t$, calculated as 
${\rm{MSD}} =(1/N)\sum_{j=1}^{N}\left|\mathbf{r}_j(t)-\mathbf{r}_j(0)\right|^{2}$. We have included data for several values of $\eta$ by keeping the system linear dimension $L$ fixed at $30$. The data set for the largest value of $\eta$ is consistent with nearly diffusive behavior from the very beginning. With the decrease of $\eta$, however, an initial part tends to be consistent with a ballistic transport. A crossover from such a rapid early displacement to a slower behavior gets delayed with further decrease of $\eta$. In Fig. \ref{fig_msd1} (b), for the same model, we plot \cite{vicsek1995novel} $v_a$, as a function of $t$, for the same set of $\eta$. We have calculated the MSD from the regime following the time when the systems reach the steady states, marked by saturation in $v_a$, corresponding to a given value of $\eta$.

\begin{figure}[h]
\centering
\includegraphics*[width=0.48\textwidth]{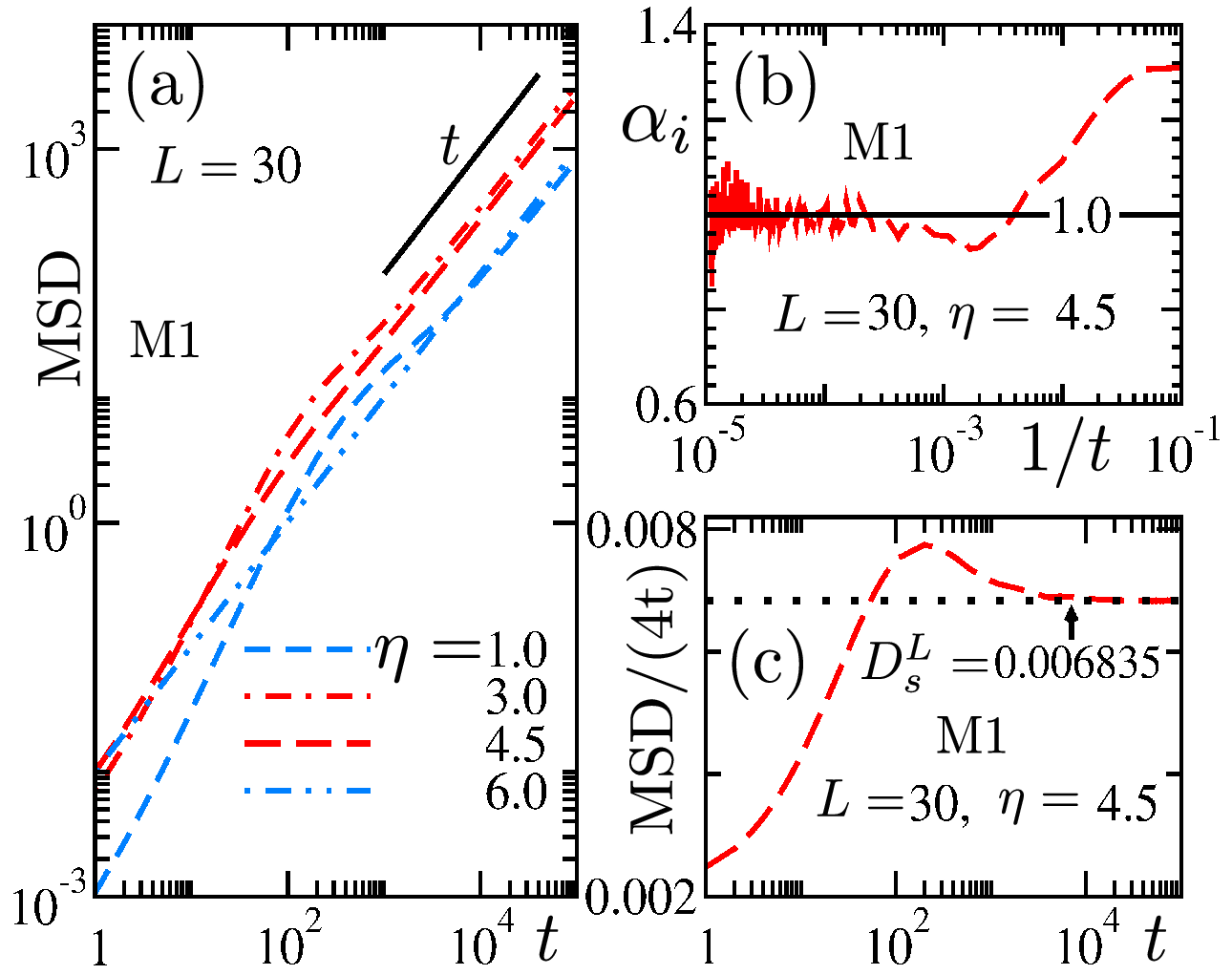}
\caption{(a) Same as Fig. \ref{fig_msd1} (a), but here MSD is calculated in the centre of mass reference frame. The solid line represents a linear (diffusive) behavior. (b) The instantaneous exponent, $\alpha_i$, obtained from a logarithmic derivative of the MSD, is plotted versus $1/t$, for the case $\eta=4.5$. The horizontal line is drawn to confirm a diffusive behavior in the late time limit. (c) For $\eta=4.5$, ${\rm{MSD}}/4t$ is plotted versus time. The dotted horizontal line demonstrates the estimation of the finite-size diffusion constant, $D_s^L$. These results are for M1.}
\label{fig_msd2}
\end{figure}

The saturations of $v_a$ at different values provide information on the order-disorder transition, implying that there can be significant nonzero centre-of-mass (CM) velocities. In such a situation, in addition to the possibility of missing out on the details of internal dynamics within the cohort, there are following further disadvantages if the displacements, as presented in Fig. \ref{fig_msd1} (a), are calculated from a fixed frame of reference: a) With the reduction of $\eta$, a crossover, from (early time) fast to (later time) slow displacements, keeps getting delayed, making the latter computationally difficult to access; b) Even though the $\eta$-dependence appears in the pre-crossover regime, as the behavior of MSD switches gradually from diffusive to a ballistic motion, the values  of the exponent --- MSD $\sim t^{\beta}$ --- changes continuously between $1$ and $2$. When $\beta$ is plotted versus $\eta$, it reveals informations similar to the phase diagram in $v_a$ versus $\eta$ plane. We, thus, calculate the MSD in the CM coordinate system. Plots from such calculations, for M1, are shown in Fig. \ref{fig_msd2} (a). It is quite evident that, despite differences in the early period, late time behavior, for each $\eta$, is linear, i.e., there exists clear diffusive transport \cite{hansen2013theory}. For a clearer confirmation of this we have calculated an instantaneous exponent \cite{huse1986corrections,amar1988monte,majumder2011diffusive}, $\alpha_i=\dv*{\ln{\rm{MSD}}}{\ln t}$, and plotted it in Fig. \ref{fig_msd2} (b), for a particular choice of $\eta$. Consistency of the data with ordinate value unity reconfirms diffusive behavior. Our adopted method to estimate the diffusion constant is demonstrated in Fig. \ref{fig_msd2} (c). There we have plotted ${\rm{MSD}}/4t$ as a function of time. The late time constant value provides $D_s$. With the anticipation of an influence from the critical point, which may bring in system-size dependence, here we introduce the symbol $D_s^L$ for this quantity.

\begin{figure}
\centering
\includegraphics*[width=0.48\textwidth]{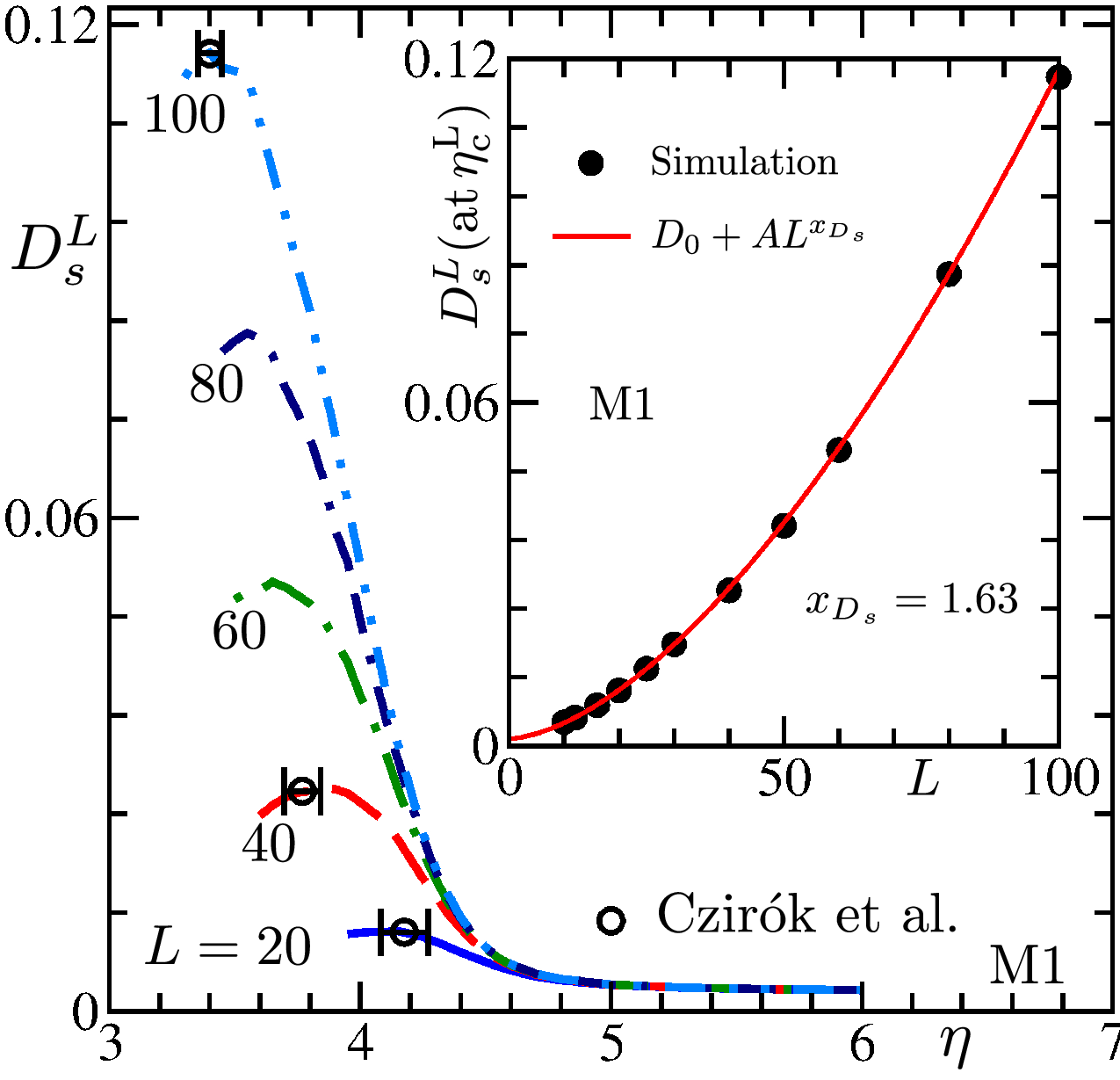}
\caption{Diffusion constant, $D_s^L$, for different system sizes, are plotted as a function of $\eta$, for the M1. The maxima locate $\eta_c^L$, the finite-size critical points. For a few system sizes the estimates of $\eta_c^L$, from a different route, by Czir\'{o}k et al. \cite{czirok1997spontaneously}, are marked by open circles. The values of $D_s^L$ at $\eta_c^L$ are plotted versus the system size in the inset. The continuous line there represents a fit of the simulation data set to the functional form mentioned inside the frame. This provides a power-law exponent $x_{D_{s}}\simeq1.6$. }
\label{fig_etacl}
\end{figure}

In the main frame of Fig. \ref{fig_etacl}, for the M1, we plot $D_s^L$, versus $\eta$, for several different $L$. Interestingly, data set for each $L$ exhibits nonmonotonic behavior, with significant finite-size effects near the peaks. In the theoretical literature of critical phenomena the location of the maximum in such a plot is considered to be a ``critical'' point for a finite system \cite{plischke1994equilibrium,das2006critical,Roy2014JCP,midya2017finite}. However, as stated above, in passive matter phase transitions, behavior of $D_s$ has not been recognized to contain a critical signature. Furthermore, it is also known that slowing down of dynamics leads to \textit{vanishing} collective diffusivity at a critical point \cite{hohenberg1977theory,tauberBook2014critical,burstyn_sengers_1980dynamical,hao_ferrell_bhatt_2005critical,chen2005universal,das2006critical}. Thus, this observation of giant enhancement of the microscopic self-diffusivity is certainly surprising! 
Recall that an order-disorder transition in $d=2$ itself is a matter of surprize \cite{toner1995,toner1998flocks}. The directionality here possesses a continuous symmetry. An inter-particle potential, irrespective of a short or long range, is absent in the 2D model. Furthermore, the dynamical rule is associated with neighbors that are not distant enough. These lead to an impression that the requirement of the MW theorem \cite{Mermin1966,Hohenberg_qlro} are not satisfied for producing a long-range order. However, flocking occurs in real systems --- in herds of sheep, schools of fish and elsewhere --- even in low dimensions.

\begin{figure}
\centering
\includegraphics*[width=0.48\textwidth]{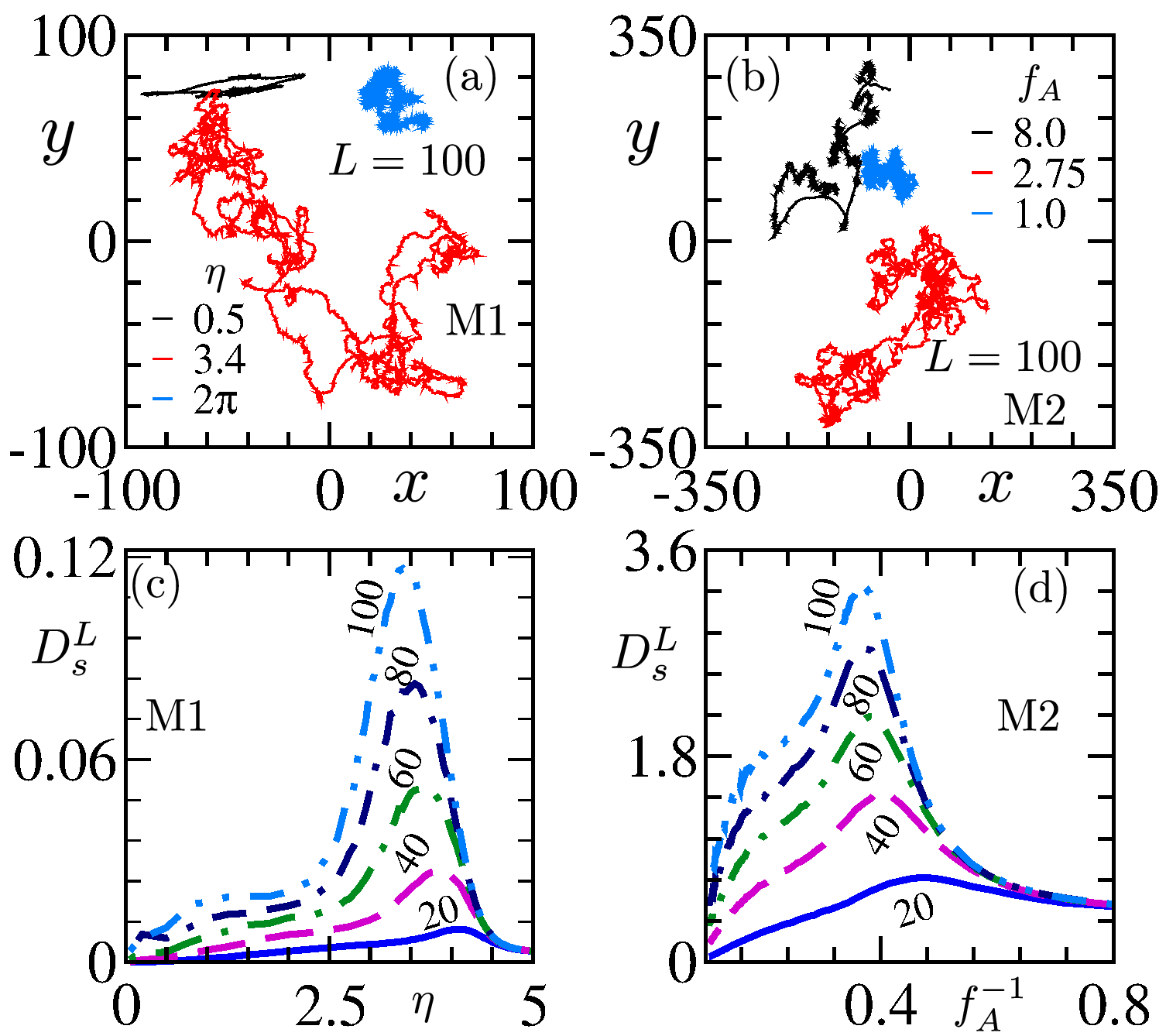}
\caption{Single-particle trajectories are shown for (a) M1 and (b) M2 for different $\eta$ and $f_A$ values. (c) $D_s^L$ versus $\eta$, for different system sizes, are plotted for the M1, over a long range of $\eta$. (d) Same as (c) but here the results are for M2, with $D_s^L$  plotted as a function of $f_A^{-1}$. For M2 we have fixed $T$ and $\rho$ at values $\epsilon/k_B$ and $0.3$, respectively, $k_B$ being the Boltzmann constant. We set integration time step for this model at $0.01\tau$, $\tau~(=\sqrt{m\sigma^2/\epsilon})$ being the LJ unit of time. We have also fixed $m$, $\sigma$, $\epsilon$, and $\gamma$ to unity.}
\label{fig_diff}
\end{figure}

The locations of the maxima in Fig. \ref{fig_etacl} match nicely with the available estimates of finite-size critical $\eta$, by Czir\'{o}k et al. \cite{czirok1997spontaneously}, via another route.  See, for comparison, the circles on or near the peaks of the data sets for $L=20$, $40$ and $100$. These values of $\eta$ will henceforth be referred to as $\eta_c^L$. 
The observed critical behavior can be quantified \cite{plischke1994equilibrium} from a finite-size scaling theory, involving $\eta_c^L$, noting that at the latter points $\xi \sim L$. 
Thus, with the objective of quantifying $D_s$, as a function of $\xi$, for M1, in the inset of Fig. \ref{fig_etacl} we plot $D_s^L$, estimated at $\eta_c^L$, with the variation of $L$. A sharp rise with a  super-linear behavior can be appreciated. The continuous line there is a fit of the simulation data set to the form $D_s^L=D_0+AL^{x_{D_{s}}}$,
$D_0$ and $A$ being constants. This captures an enhancement $D_s\sim\xi^{x_{D_s}}$, over the considered range of $L$, with a strong exponent $x_{D_{s}}\simeq 1.6$.

The Berezinskii-Kosterlitz-Thouless transition \cite{Kosterlitz1973,Kosterlitz1974} is related to quasi-long-range order in 2D XY model and other systems. This respects the prohibition stated in the MW theorem, on the spontaneous breaking of continuous symmetry in absence of a long range interactions. However, even though the symmetry in the velocity direction is continuous in our study, as for the spins in the XY model, the mobilities of the particles in the present study, particularly when diffusion is so strong at $\eta_c$, make the situation different from the XY case in a significant way. Owing to this, an effective interaction may originate \cite{toner1995,toner1998flocks}, having range much longer than the Vicsek radius $\mathcal{R}_{\rm{M1}}$. If the range becomes significant, the constraint imposed by the MW theorem should get evaded. However, the question is concerning the existence of order and thus, enhanced diffusion throughout the region below the $\eta_c$.  

In Fig. \ref{fig_diff} (a) we show single-particle trajectories for M1, corresponding to three different values of $\eta$ chosen from the ordered, disordered and the critical regions. It can be noted how at the critical value, i.e., at $\eta=3.4$, the particle is swirling around a much larger space. Fig. \ref{fig_diff} (b) contains similar plots but for M2. These convey that in the ordered region the diffusion is not as enhanced, as opposed to that near criticality.
However, for enhanced diffusion to be the reason behind ordering, bypassing the MW theorem, strong transport must exist throughout the full ordered region of the phase diagram. The requirement, however, can be satisfied if $D_s$ exhibits exceptional finite-size effects not only in the vicinity of the critical points, but also in the entire ordered region.

\begin{figure}
\centering
\includegraphics*[width=0.48\textwidth]{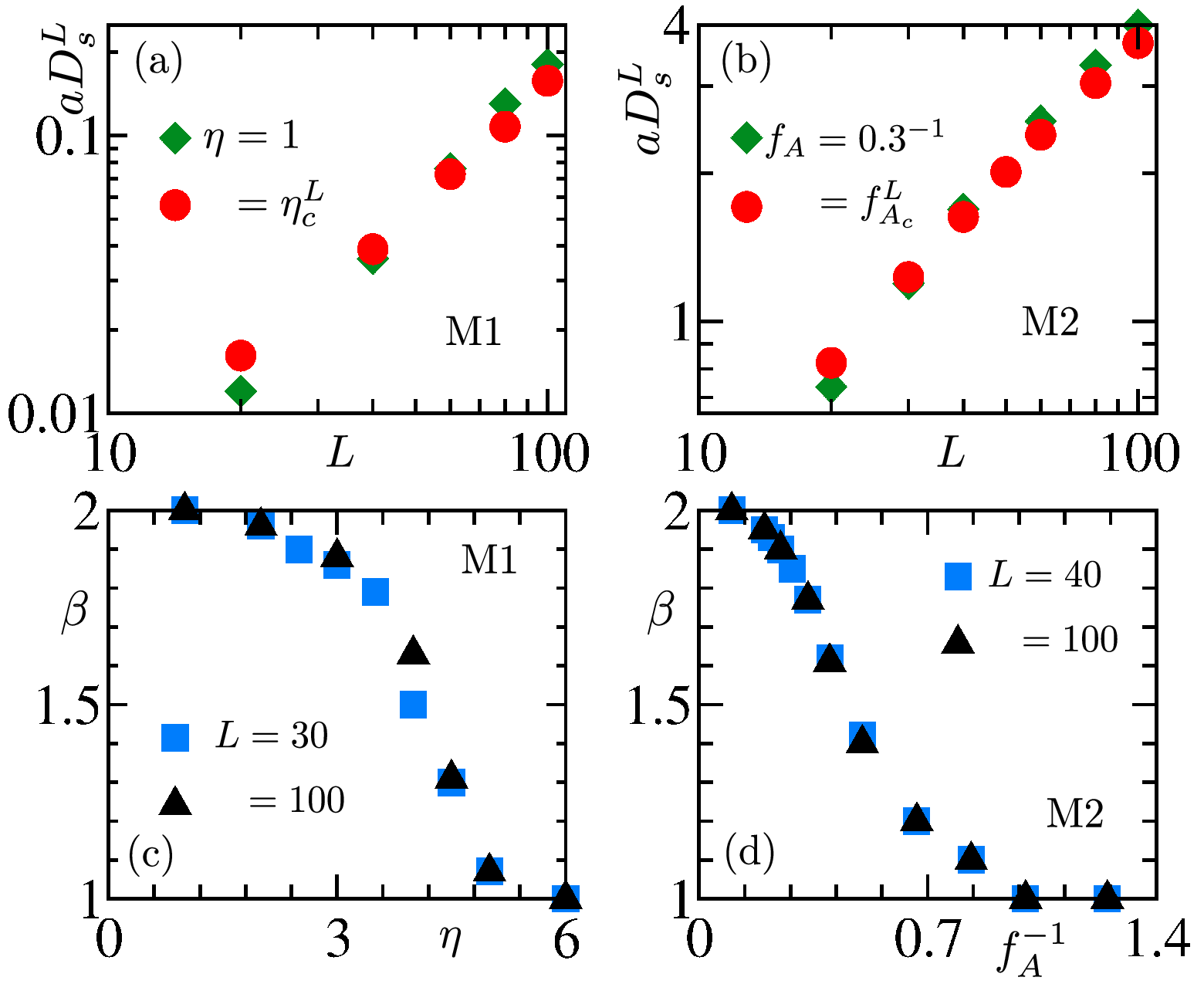}
\caption{(a) Plots of $D_s^L$ versus $L$,  for the M1, at $\eta=1$ and $\eta_c^L$.  The ordinate of the data sets are scaled by constant factors ($a$) to obtain overlap. (b) Same as (a) but here the results are for M2, at $f_A=0.3^{-1}$ and $f_{A_c}^L$. (c) Plots of $\beta$ are shown from two system sizes, for (c) M1 (vs $\eta$) and (d) M2 (vs $f_A^{-1}$).}
\label{fig_inst_diff}
\end{figure}

In Fig. \ref{fig_diff} (c) we present data for M1, from same system sizes as in the main frame of Fig. \ref{fig_etacl}, over a much fuller range. Indeed, very strong finite-size effects appear almost in the entire ordered region. This is unusual in standard critical phenomena \cite{fisherRevModPhys.46.597,stanleyBook,onukiBook2002} where data from different system sizes, for a quantity, merge with each other within a short distance away from the critical point, on either side. This merger can be appreciated here as well, but only for $\eta>\eta_c$. Our observation of finite-size behavior, in the ordered region, on the other hand, is \textit{peculiar}. Nevertheless, this complies with the above stated requirement for long-range order. In an effective sense, this implies that the particles interact over distances that are longer than the assigned range of active interaction. The data suggest, except for the zero noise case, diffusion is very strong in thermodynamic large systems. This overall picture, hitherto unreported, we believe, is consistent with the formulation of the Toner-Tu theory \cite{toner1995,toner1998flocks}. In Fig. \ref{fig_diff} (d) we present similar data from M2. The features remain same. It appears, interestingly, that there exist points of inflection, irrespective of the model. This is an artifact arising out of vanishing diffusivity at no noise scenario and strong size effects elsewhere having a peak at criticality. In Fig. \ref{fig_inst_diff} (a) and (b) we compare the finite-size effects in the ordered regions, for both the models, with the corresponding critical pictures. It appears that the strength of the effects remain similar, even when the state points are chosen quite deep inside the ordered region.
As discussed earlier, if transport phenomena is studied within the framework of a fixed frame of references, the crossover to a slower MSD keeps getting shifted to times that may not be accessible within the simulations time scales. One way, in that case, to extract the dependence of transport on $\eta$ or $f_A$ is to estimate the power-law exponent $\beta$ before the final crossover to a slower MSD. This quantity is plotted in Fig. \ref{fig_inst_diff} (c) and (d) for M1 and M2. In both the cases a non-remarkable continuous variation is seen. Furthermore, there exists no finite-size effects.

We have studied transport phenomena across flocking transitions in two different active matter models. In the centre of mass coordinate system particles exhibit diffusive motion. There exist strong finite-size effects in $D_s$, the self-diffusion constant, at the critical points. This implies robust critical enhancements in the thermodynamic system-size limit. Our understanding of transport phenomena in equilibrium phase transitions is in deviation with this. Furthermore, when these results are combined with the behavior of shear viscosity \cite{Arash2019}, violation of the Stokes-Einstein-Sutherland relation \cite{hansen2013theory} becomes apparent. Intriguingly, the finite-size effects continue to remain strong throughout the ordered region. This overall counterintuitive dynamical picture has important consequences. The enhanced diffusion, we believe, invoke long-range effective interaction \cite{toner1995,toner1998flocks}. This is important for establishing order in the system by evading the prohibition stated in the Mermin-Wagner theorem \cite{Mermin1966,Hohenberg_qlro} for lower dimensional systems.
On this aspect, the results from both the models bear strong qualitative agreement with the basic expectation of the Toner-Tu theory \cite{toner1995,toner1998flocks}.
This generic hydrodynamic model provides a platform for understanding similar phenomena in a large class of systems. We, however, do not expect quantitative agreement with the theory given that universality is weak in active matter.

The authors acknowledge the computational facility and financial support from JNCASR.

The data will be made available upon reasonable request to the authors.

\bibliography{references}

\end{document}